\documentclass[a4paper]{article}

\usepackage{INTERSPEECH2022}
\usepackage{url}
\usepackage{cite}
\usepackage{multirow}

\DeclareMathOperator{\diag}{diag}

\usepackage[linesnumbered,ruled,vlined]{algorithm2e}

\newcommand{\ReduceSpaceUnderFigure}{\vspace{-10pt}}
\newcommand{\ReduceSpaceUnderTable}{\vspace{-8pt}}

\title{SpecGrad: Diffusion Probabilistic Model based Neural Vocoder with\\
Adaptive Noise Spectral Shaping}
\name{Yuma Koizumi$^1$, Heiga Zen$^1$, Kohei Yatabe$^2$, Nanxin Chen$^1$, Michiel Bacchiani$^1$}
\address{$^1$Google Research, $^2$Tokyo University of Agriculture and Technology}
\email{\{koizumiyuma,heigazen,nanxinchen,michiel\}@google.com}

\begin{document}

\maketitle

\begin{abstract}
Neural vocoder using denoising diffusion probabilistic model (DDPM) has been improved by adaptation of the diffusion noise distribution to given acoustic features.
In this study, we propose \textit{SpecGrad} that adapts the diffusion noise so that its time-varying spectral envelope becomes close to the conditioning log-mel spectrogram.
This adaptation by time-varying filtering improves the sound quality especially in the high-frequency bands.
It is processed in the time-frequency domain to keep the computational cost almost the same as the conventional DDPM-based neural vocoders.
Experimental results showed that SpecGrad generates higher-fidelity speech waveform than conventional DDPM-based neural vocoders in both analysis-synthesis and speech enhancement scenarios.
Audio demos are available at \url{wavegrad.github.io/specgrad/}.
\end{abstract}

\section{Introduction}
\label{sec:intro}

Neural vocoders~\cite{sample_rnn,tamamori2017speaker,waveglow,waveflow} are neural networks that generate a speech waveform given acoustic features.
They are indispensable building blocks of recent speech applications.  For example, they are used as a backend model of text-to-speech (TTS)~\cite{sample_rnn,tacotron2} and speech enhancement (SE)~\cite{Maiti_waspaa_2019,Maiti_icassp_2020,Su_2020,Su_2021,voice_filxer}.
A challenge in neural vocoder research is to generate a high-fidelity speech waveform with low computational costs.
Autoregressive models~\cite{wavenet,sample_rnn,wavernn} have revolutionized the quality of output signals, yet the nature of the models requires a large number of sequential operations for generation.
To speed up the inference, various non-autoregressive models have been proposed such as generative adversarial networks~\cite{Donahue_2019,Kong_2020,melgan,Yamamoto_2020,Yang_2021}, flow-based models~\cite{waveglow,waveflow}, and signal processing-inspired models~\cite{juvela-taslp-2019,neural_source_filter,period_net}.

\begin{figure}[t]
\vspace{-1pt}
  \centering
\includegraphics[width=\linewidth,clip]{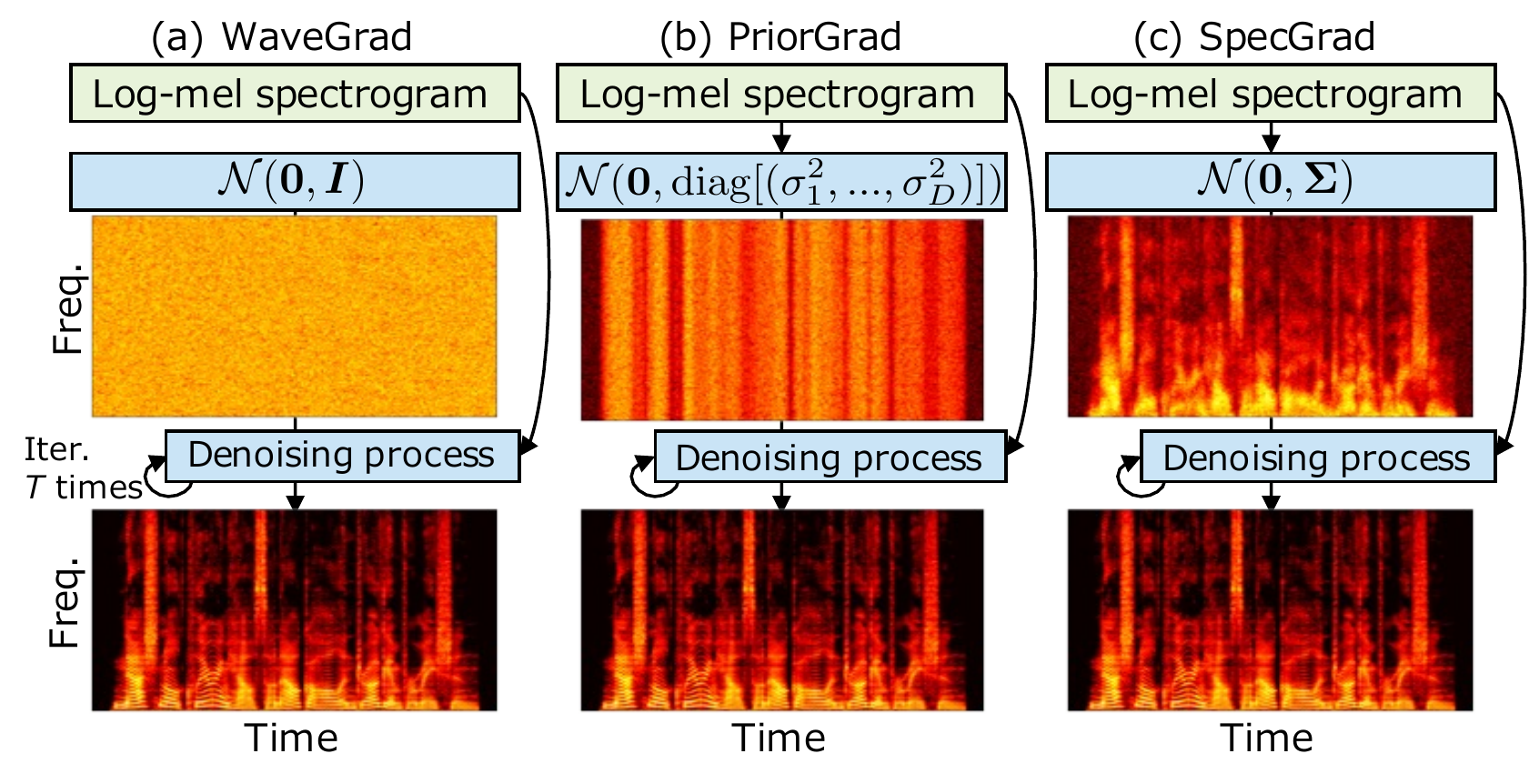} 
  \vspace{-20pt}
  \caption{(a) WaveGrad~\cite{wavegrad}, (b) PriorGrad with a diagonal covariance matrix~\cite{priorgrad}, and (c) Proposed SpecGrad.
  While (a) samples noise from the standard Gaussian, (b) and (c) sample noise from Gaussian with covariance matrix calculated from the conditioning log-mel spectrogram.
  Proposed SpecGrad in (c) adapts noise to both signal power and spectral envelope.}
  \label{fig:concept}
  \ReduceSpaceUnderFigure
\end{figure}

Among non-autoregressive models, denoising diffusion probabilistic models (DDPMs)~\cite{Ho_2020} have recently gained increased attention due to its quality and the controllable computational cost~\cite{wavegrad,diffwave,priorgrad,Bddm022,InferGrad2022,okamoto2021,Goel_2022}.
DDPMs convert a random signal into a speech waveform by the iterative sampling process called \textit{denoising process} as illustrated in Fig.~\ref{fig:concept}~(a).
Since it iteratively refines a waveform, DDPMs have a tradeoff between the output quality and computational costs~\cite{wavegrad}, i.e., many iterations are necessary for obtaining a high-fidelity waveform.
To reduce the number of iterations while maintaining the quality, the conventional studies have proposed a proper inference noise schedule~\cite{Bddm022,InferGrad2022} and/or network architecture~\cite{Goel_2022,okamoto2021}.

PriorGrad~\cite{priorgrad} provided a new approach to DDPM-based neural vocoders by considering the prior distribution on the acoustic model~\cite{grad_tts}.
It adapts diffusion noises based on the conditioning log-mel spectrogram as illustrated in Fig.~\ref{fig:concept}~(b).
More specifically, the diffusion distribution is Gaussian with the diagonal covariance matrix whose diagonal entries are frame-wise energies of the mel-spectrogram.
This proposal can be regarded as scaling of noise schedule for each sample point because a diagonal covariance matrix represents the power of a waveform in the time domain.
Introduction of this well-known relation in signal processing to DDPMs is one of the important contributions of PriorGrad.
The success of PriorGrad suggests that DDPM-based neural vocoders can be improved further by incorporating more knowledge from signal processing. 

In this study, we propose \textit{SpecGrad} that adapts spectral envelope of diffusion noise to the conditioning log-mel spectrogram as illustrated in Fig.~\ref{fig:concept}~(c).
We begin our discussion with the decomposed covariance matrix that is used in both diffusion noise generation and cost function of DDPMs.
Then, we design a covariance matrix to manipulate the spectral envelope of diffusion noise, which is realized by time-varying filtering in the time-frequency (T-F) domain.
We conducted objective and subjective experiments to show that SpecGrad generates a waveform whose quality is better than WaveGrad~\cite{wavegrad} and PriorGrad~\cite{priorgrad} on both analysis-synthesis and SE tasks.

% -------------------------------------------------------------------------
\section{Conventional methods}
\label{sec:diffusion}

\subsection{DDPM-based neural vocoder}
\label{sec:wavegrad}

Let a $D$-point speech waveform $\bm{x}_0 \in \mathbb{R}^D$ be generated from a conditioning log-mel spectrogram $\bm{c} = (\bm{c}_1,...,\bm{c}_K) \in \mathbb{R}^{FK}$, where $\bm{c}_k \in \mathbb{R}^{F}$ is an $F$-point log-mel spectrum at $k$th time frame, and $K$ is the number of time frames.
Our goal is to find a probability density function (PDF) of $\bm{x}_0$ as $q(\bm{x}_0 \mid \bm{c})$. 

A DDPM-based neural vocoder is a latent variable model based on a Markov chain of $\bm{x}_{t} \in \mathbb{R}^D$ with learned Gaussian transitions, starting from $q(\bm{x}_T) = \mathcal{N}(\bm{0}, \bm{I})$, defined as
\vspace{-5pt}
\begin{align}
    q(\bm{x}_0 \mid \bm{c}) = \int_{\mathbb{R}^{DT}} \!\!\!\!\!
    q(\bm{x}_T) \prod_{t=1}^T q(\bm{x}_{t-1}  \mid \bm{x}_{t}, \bm{c}) \, \mathrm{d} \bm{x}_{1} \cdots \mathrm{d} \bm{x}_{T}.    
\end{align}
By modeling $q(\bm{x}_{t-1} \mid \bm{x}_{t}, \bm{c})$, $\bm{x}_0$ can be obtained from $\bm{x}_T$ via recursive sampling of $\bm{x}_{t-1}$ from $\bm{x}_t$.

In DDPMs, $\bm{x}_{t-1}$ is generated by the \textit{diffusion process} that gradually adds Gaussian noise to the waveform according to a noise schedule $\{\beta_1,...,\beta_T\}$ given by $p(\bm{x}_t \mid \bm{x}_{t-1}) = \mathcal{N}\left(\sqrt{1 - \beta_t}\bm{x}_{t-1}, \beta_t \bm{I} \right)$.
This formulation enables us to sample $\bm{x}_t$ at an arbitrary timestep $t$ in a closed form as 
\begin{align}
    \bm{x}_t = \sqrt{\bar{\alpha}_{t}} \bm{x}_{0} + \sqrt{1 - \bar{\alpha}_{t}}\bm{\epsilon},
    \label{eq:diffusion}
\end{align}
where $\alpha_t = 1 - \beta_t$, $\bar{\alpha}_{t} = \prod_{s=1}^t \alpha_s$, and $\bm{\epsilon} \sim \mathcal{N}(\bm{0}, \bm{I})$.
As proposed by Ho \textit{et al.}~\cite{Ho_2020}, DDPM-based neural vocoders use a deep neural network (DNN) $\mathcal{F}$ with parameter $\theta$ for predicting $\bm{\epsilon}$ from $\bm{x}_t$ as $\bm{\epsilon} = \mathcal{F}_{\theta}(\bm{x}_t, \bm{c}, \beta_t)$.
Then, if $\beta_t$ is small enough, $q(\bm{x}_{t-1} \mid \bm{x}_{t}, \bm{c})$ can be given by $\mathcal{N}(\bm{\mu}_{t}, \gamma_t \bm{I})$, where
\begin{align}
    \bm{\mu}_{t} = \frac{1}{\sqrt{\alpha_t}} \left( \bm{x}_t - \frac{1 - \alpha_t}{\sqrt{1 - \bar{\alpha}_t}} \mathcal{F}_{\theta}(\bm{x}_t, \bm{c}, \beta_t) \right), \label{eq:decode_mu}
\end{align}
and $\gamma_t = \frac{1 - \bar{\alpha}_{t-1}}{1 - \bar{\alpha}_{t}}\beta_t$.
The DNN $\mathcal{F}$ can be trained by maximizing the evidence lower bound (ELBO), though most of DDPM-based neural vocoders use a simplified loss function for the training; for example, WaveGrad~\cite{wavegrad} used the $\ell_1$ norm as 
\begin{align}
    \mathcal{L}^{\mbox{\tiny WG}} = \left\lVert \bm{\epsilon} - \mathcal{F}_{\theta}(\bm{x}_t, \bm{c}, \beta_t) \right\rVert _{1}
    , \label{eq:wavegrad_loss}
\end{align}
where $\lVert \cdot \rVert_p$ denotes the $\ell_p$ norm.

\vspace{-2pt}
\subsection{PriorGrad}
\label{sec:priorgrad}

Lee \textit{et al}.\ proposed PriorGrad~\cite{priorgrad} by introducing an adaptive prior $\mathcal{N}(\bm{0}, \bm{\Sigma})$, where $\bm{\Sigma}$ is computed from $\bm{c}$.
Compared to the conventional DDPM-based neural vocoders, PriorGrad is different in two points:
(i) $\bm{\epsilon}$ is sampled from $\mathcal{N}(\bm{0}, \bm{\Sigma})$ for all diffusion steps, and
(ii) the Mahalanobis distance according to $\bm{\Sigma}$,
\begin{align}
    \mathcal{L}^{\mbox{\tiny PG}} = (\bm{\epsilon} - \mathcal{F}_{\theta}(\bm{x}_t, \bm{c}, \beta_t))^{\top} \bm{\Sigma}^{-1}  (\bm{\epsilon} - \mathcal{F}_{\theta}(\bm{x}_t, \bm{c}, \beta_t)), \label{eq:priorgrad_loss}
\end{align}
is used for the loss function, where ${}^{\top}$ is the transpose.
In their experiments, a specific form of the covariance matrix was given by $\bm{\Sigma} = \diag[(\sigma^2_1, \sigma^2_2, ..., \sigma^2_D)]$, where $\sigma^2_d$ denotes the signal power at $d$th sample calculated by interpolating the normalized frame-wise energy of $\bm{c}_k$~\cite{priorgrad}, and $\diag$ constructs the diagonal matrix whose diagonal entries are those of the input vector.

% -------------------------------------------------------------------------
\section{Proposed method}
\label{sec:proposed}

\subsection{SpecGrad}
\label{sec:prop_basic}

The performance of PriorGrad is determined by the covariance matrix $\bm{\Sigma}$.
As Lee \textit{et al}.\ showed, ELBO of PriorGrad becomes small when $\bm{\Sigma}$ is close to the covariance of $\bm{x}_0$~\cite{priorgrad}.
One way to obtain such $\bm{\Sigma}$ is to make the amplitude spectrum of $\bm{\epsilon}$ similar to that of $\bm{x}_0$.
In this paper, we propose \textit{SpecGrad} by incorporating the information of spectral envelope into $\bm{\Sigma}$.

Since $\bm{\Sigma}$ is positive semi-definite, it can be decomposed as $\bm{\Sigma} = \bm{L} \bm{L}^{\top\!}$. 
Then, sampling from $\mathcal{N}(\bm{0}, \bm{\Sigma})$ can be written as $\bm{\epsilon} = \bm{L} \tilde{\bm{\epsilon}}$ using $\tilde{\bm{\epsilon}} \sim \mathcal{N}(\bm{0}, \bm{I})$, and Eq.~\eqref{eq:priorgrad_loss} can be rewritten as
\begin{align}
    \mathcal{L}^{\mbox{\tiny SG}} &= \left\lVert \bm{L}^{-1} (\bm{\epsilon} - \mathcal{F}_{\theta}(\bm{x}_t, \bm{c}, \beta_t)) \right\rVert_2^2. \label{eq:our_loss}
\end{align}
Thus, our interest is to design $\bm{L} \in \mathbb{R}^{D \times D}$ so that a high-fidelity waveform can be generated with low computational costs.

Some desired properties of $\bm{L}$ are as follows.
First, the amplitude spectrum of $\bm{\epsilon}$ $(= \bm{L} \tilde{\bm{\epsilon}})$ approximates that of $\bm{x}_0$.
This property is required to lower ELBO.
Second, multiplication of $\bm{L}$ and $\bm{L}^{-1\!}$ should be efficiently computed.
Their computation directly impacts the total cost of training because $\bm{L}$ and $\bm{L}^{-1\!}$ are repeatedly applied in the training and depend on the training sample $\bm{x}_0$ due to the first property.

To meet the requirements given in the previous paragraph, we propose to apply a time-varying filter in the T-F domain.
Let the short-time Fourier transform (STFT) be represented by an $NK \times D$ matrix $\bm{G}$, where $N$ is the window size.
We consider the following time-varying filter in the T-F domain:
\begin{equation}
    \bm{L} = \bm{G}^{+}\bm{M}\bm{G},
    \label{eq:defPropL}
\end{equation}
where $\bm{M} = \diag[(m_{1,1}, \ldots, m_{N,K})] \in \mathbb{C}^{NK \times NK}$ is the diagonal matrix that defines the filter, $m_{n,k} \neq 0$ is a coefficient multiplied to the $(n,k)$th T-F bin, and $\bm{G}^{+\!}$ is the matrix representation of the inverse STFT (iSTFT) using a dual window.%
\footnote{%
We define STFT and iSTFT so that they satisfy $\bm{G}^{+}\bm{G} = \bm{I}$ and $\bm{G}^{+}\bm{M}\bm{G}\in\mathbb{R}^{D \times D}$.
This is realized by (i) using a pair of windows that fulfills the perfect reconstruction condition and (ii) preserving the conjugate symmetry of spectra within $\bm{G}^{+}\bm{M}$.}
% Note that these matrices will not be constructed but implemented as functions.}
This representation allows us to recast the problem of designing $\bm{L}$ to a filter design problem.
We propose to design $\bm{M}$ so that $\bm{\epsilon}$ $(= \bm{L} \tilde{\bm{\epsilon}})$ approximates the spectral envelope of $\bm{x}_0$.
We also propose to approximate $\bm{L}^{-1\!}$ as $\bm{L}^{-1} \approx \bm{G}^{+}\bm{M}^{-1}\bm{G}$.

Since $\bm{M}$ and $\bm{M}^{-1\!}$ are diagonal, both $\bm{L}$ and the approximate $\bm{L}^{-1\!}$ can be applied with $O(K N \log N)$ operations using a fast Fourier transform (FFT) algorithm.
In addition, $K$ FFTs can be computed in parallel.
Therefore, Eq.~\eqref{eq:defPropL} provides a good compromise between the flexibility and computational cost.

\vspace{-2pt}
% -------------------------------------------------------------------------
\subsection{Implementation}
\label{sec:prop_implementation}

\begin{algorithm}[t]
\DontPrintSemicolon
\caption{Training of SpecGrad.}
\label{algo:training}
\SetKwFunction{FMain}{TrainOneStep}
\SetKwFunction{FSub}{SampleNoise}
\SetKwProg{Fn}{Function}{:}{\KwRet}
\Fn{\FMain{$\bm{x}_0$, $\bm{c}$, $\bm{M}$, $\beta_t$}}{
    $\bm{\epsilon} \gets \FSub(\bm{M})$\\
    $\bm{x}_t \gets \sqrt{\bar{\alpha}_{t}} \bm{x}_{0} + \sqrt{1 - \bar{\alpha}_{t}}\bm{\epsilon}$\\
    $\mathcal{L} \gets \left\lVert 
    \bm{G}^{+}\bm{M}^{-1} \bm{G} (\bm{\epsilon} - \mathcal{F}_{\theta}(\bm{x}_t, \bm{c}, \beta_t)))  \right\rVert_2^2$ \\
    Update the model parameter $\theta$ based on $\nabla_{\theta} \mathcal{L}$
}
\Fn{\FSub{$\bm{M}$}}{
    Sample $\tilde{\bm{\epsilon}} \sim \mathcal{N}(\bm{0}, \bm{I})$\\
    \KwRet $\bm{G}^{+}\bm{M}\bm{G} \tilde{\bm{\epsilon}}$
}
\end{algorithm}
\begin{algorithm}[t]
\DontPrintSemicolon
\caption{Inference of SpecGrad.}
\label{algo:sampling}
\SetKwFunction{FMain}{Sampling}
\SetKwFunction{FSub}{SampleNoise}
\SetKwProg{Fn}{Function}{:}{\KwRet}
\Fn{\FMain{$\bm{c}$, $\bm{M}$, $\beta_{1},\ldots,\beta_{T}$}}{
    $\bm{x}_T \gets \FSub(\bm{M})$\\
    \For{$t = T$ to $1$}{
        $\hat{\bm{\epsilon}} \gets \mathcal{F}_{\theta}(\bm{x}_t, \bm{c}, \beta_t)$\\
        $\bm{x}_{t-1} \gets \frac{1}{\sqrt{\alpha_t}}\bigl( \bm{x}_t - \frac{\beta_t}{\sqrt{1 - \bar{\alpha}_t}} \hat{\bm{\epsilon}} \bigr)$\\
        \If{$t > 1$}{
        $\bm{x}_{t-1} \gets \bm{x}_{t-1} + \gamma_t \cdot \FSub(\bm{M})$
        }
    }
    \KwRet $\bm{x}_{0}$
}
\end{algorithm}

Pseudocodes of the training and inference of the proposed method are shown in {\bf Algorithm \ref{algo:training}} and {\bf \ref{algo:sampling}}, respectively.
The differences from the vanilla DDPM-based neural vocoders~\cite{wavegrad,diffwave} and PriorGrad are (i) the diffusion noise sampling $\bm{G}^{+}\bm{M}\bm{G} \tilde{\bm{\epsilon}}$ and (ii) the loss function in Eq.~\eqref{eq:our_loss}.
Thus, the proposed method can coexist with the noise schedules and/or network architectures in the literature~\cite{Bddm022,InferGrad2022,Goel_2022,okamoto2021}.
Although SpecGrad is slightly more costly than WaveGrad, the inference speed of SpecGrad and WaveGrad is almost the same because the forward propagation of a typical DNN used in the neural vocoders is significantly more expensive than STFT.

For computation of the T-F domain filter $\bm{M}$, we estimate the spectral envelope via cepstrum as follows.
First, pseudoinverse of the mel-compression matrix is applied to $\bm{c}$ for computing the corresponding power spectrogram.
Then, the $r$th order lifter is applied to compute the spectral envelope for each time frame.
As with PriorGrad, to ensure numerical stability during training~\cite{priorgrad}, we add $0.01$ to the estimated envelope.
The coefficients $m_{1,k}, \ldots, m_{N,k}$ for the $k$th time frame are computed from the obtained envelope with minimum phase response.

Note that any other method can be a choice for constructing the filter $\bm{M}$.
The above choice was due to our preliminary investigation.
Directly using the spectrogram did not provide a satisfactory result, and thus we apply envelope estimation.
We chose the cepstrum-based spectral envelope estimation method because its implementation is simpler than the other methods.
These arguments are based on our informal experiments, and detailed study of the filter design is left as a future work.

% -------------------------------------------------------------------------
\section{Experiment}
\label{sec:experiment}
In this paper, we compared SpecGrad with WaveGrad~\cite{wavegrad} and PriorGrad~\cite{priorgrad} by both objective and subjective experiments.
We also evaluated their performance as a backend module of speech enhancement.
Since WaveGrad has been compared with an autoregressive model, WaveRNN~\cite{wavernn}, and several non-autoregressive models~\cite{melgan,Yamamoto_2020,Yang_2021} in the literature, we focus on the DDPM-based neural vocoders with different diffusion PDFs.
Audio demos are available in our demo page.\footnote{
\url{wavegrad.github.io/specgrad/}
}

\vspace{-2pt}
\subsection{Experimental setup}
\label{sec:exp_setup}

\begin{table}[ttt]
\caption{Inference noise schedules used in experiments.}
\vspace{-6pt}
\label{tab:noise_schedule}
\centering
\begin{tabular}{ c | c  }
\toprule
\textbf{Name} & \textbf{Schedule}\\	
\midrule 
WG-3 & \textsc{[3e-4, 6e-2, 9e-1]}\\ 
WG-6 & \textsc{[7e-6, 1.4e-4, 2.1e-3, 2.8e-2, 3.5e-1, 7e-1]}\\ 
PG-6 & \textsc{[1e-4, 1e-3, 1e-2, 5e-2, 2e-1, 5e-1]}\\ 
WG-50 & \texttt{linspace(1e-4, 0.05, 50)}\\ 
\bottomrule
\end{tabular}
\ReduceSpaceUnderTable
\end{table}

\vspace{2pt}
\noindent
\textbf{Dataset:} We trained the models using a proprietary speech dataset consisted of 184 hours of high quality US English speech spoken by 11 female and 10 male speakers.
For evaluation, we used 1,000 holdout samples of US English speech spoken by the same 21 speakers as the training dataset.
The signals were downsampled to 24 kHz, and then 128-dimensional log-mel spectrograms (50 ms Hann window, 12.5 ms frame shift, 2048-point FFT, and 20 Hz and 12 kHz lower and upper frequency cutoffs, respectively) were extracted as $\bm{c}$.

\vspace{2pt}
\noindent
\textbf{Model and training setup:} To evaluate the performance difference due to the difference in diffusion PDF, we used the same network architecture and noise schedule for all three methods.
We used the ``WaveGrad Base model~\cite{wavegrad}'' having 15M parameters.
We trained all models using 128 Google TPU v3 cores with a global batch size of 512.
To accelerate training, we randomly picked 120 frames (1.5 seconds, $D=36,000$ samples) as input.
We trained all models for 1M steps (around 3 days) with the optimizer setting same as that of WaveGrad~\cite{wavegrad}.

We tested two training noise schedules and several inference noise schedules listed in Table~\ref{tab:noise_schedule}, which were used in the papers of WaveGrad~\cite{wavegrad} and PriorGrad~\cite{priorgrad}.
One is the same as WaveGrad; training schedule was \texttt{linspace(1e-6, 1e-2, 1000)},%
\footnote{\texttt{linspace} is the function that returns evenly spaced real numbers over the interval specified by the first two arguments.}
and inference schedules were WG-3, WG-6, and WG-50.
The other one is the same as PriorGrad;
training schedule was \texttt{linspace(1e-4, 0.05, 50)} and inference schedules were PG-6 and WG-50.

For PriorGrad and SpecGrad, we used the generalized energy distance (GED)~\cite{ged_loss}, which was used in the first version of the PriorGrad paper,%
\footnote{\url{https://arxiv.org/pdf/2106.06406v1.pdf}}
as an auxiliary loss function.
The weight for GED was 0.01 which is the same as that in the PriorGrad paper.
The lifter order was $r=24$.
For STFT, we used the same settings as the log-mel spectrogram calculation.

\vspace{2pt}
\noindent
\textbf{Metrics:} 
To evaluate subjective quality, we rated speech naturalness on a 5-point mean opinion score (MOS) scale (1:~Bad, 2:~Poor, 3:~Fair, 4:~Good, 5:~Excellent) with rating increments of 0.5. 
% Subjects were asked to rate the naturalness of each stimulus after listening to it. 
Test stimuli were randomly chosen and presented to subjects in isolation, i.e., each stimulus was evaluated by one subject.
Each subject was allowed to evaluate up to six stimuli.
The subjects were paid native English speakers living in United States.
They were requested to use headphones in a quiet room.

For objective evaluation, we used quality prediction for generative neural speech codecs (WAPR-Q)~\cite{warp_q} which correlates with MOS of output sounds of neural vocoders.
Since the default parameters of WARP-Q were for signals sampled at 16 kHz, we changed the cut-off frequency to 10 kHz and the number of mel-frequency cpestrum coefficients (MFCCs) to 24.

\vspace{-2pt}
\subsection{Results}
\label{sec:result}

\begin{table}[ttt]
\caption{Mean opinion scores (MOS) and WARP-Q scores with their 95\% confidence intervals. All models were trained by using training noise schedule of WaveGrad~\cite{wavegrad}. GT means ground-truth.}
\vspace{-6pt}
\label{tab:wg_schedule_result}
\centering
\begin{tabular}{ c c | c c }
\toprule
\textbf{Method} & \textbf{Schedule} & \textbf{MOS ($\uparrow$)} & \textbf{WARP-Q ($\downarrow$)}\\	
\midrule
WaveGrad & & $3.56 \pm 0.08$ & $1.54 \pm 0.010$\\ 
PriorGrad & WG-3 &  $3.45 \pm 0.08$ & $1.33 \pm 0.008$\\	
SpecGrad & &  $\bm{3.88} \pm \bm{0.07}$ & $\bm{1.22} \pm \bm{0.007}$\\	
\midrule
WaveGrad & & $4.10 \pm 0.06$ & $1.38 \pm 0.009$\\ 
PriorGrad & WG-6 &  $4.01 \pm 0.07$ & $\bm{1.12} \pm \bm{0.007}$\\	
SpecGrad & &  $\bm{4.25} \pm \bm{0.06}$ & $\bm{1.12} \pm \bm{0.007}$\\	
\midrule
WaveGrad & & $4.30 \pm 0.06$ & $1.28 \pm 0.008$\\ 
PriorGrad & WG-50 &  $4.32 \pm 0.06$ & $1.11 \pm 0.007$\\	
SpecGrad & &  $\bm{4.39} \pm \bm{0.06}$ & $\bm{1.09} \pm \bm{0.006}$\\	
\midrule
GT & --- &  $4.47 \pm 0.05$ & -\\	
\bottomrule
\end{tabular}
\ReduceSpaceUnderTable
\end{table}

\begin{table}[ttt]
\caption{Mean opinion scores (MOS) and WARP-Q scores with their 95\% confidence intervals. All models were trained by using training noise schedule of PriorGrad~\cite{priorgrad}.}
\vspace{-6pt}
\label{tab:pg_schedule_result}
\centering
\begin{tabular}{ c c | c c }
\toprule
\textbf{Method} & \textbf{Schedule} & \textbf{MOS ($\uparrow$)} & \textbf{WARP-Q ($\downarrow$)}\\	
\midrule
WaveGrad & & $4.14 \pm 0.06$ & $1.32 \pm 0.009$\\ 
PriorGrad & PG-6 &  $4.02 \pm 0.06$ & $1.10 \pm 0.007$\\	
SpecGrad & &  $\bm{4.31} \pm \bm{0.05}$ & $\bm{1.05} \pm \bm{0.006}$\\	
\midrule
WaveGrad & & $4.29 \pm 0.05$ & $1.31 \pm 0.008$\\ 
PriorGrad & WG-50 &  $4.29 \pm 0.05$ & $1.08 \pm 0.007$\\	
SpecGrad & &  $\bm{4.40} \pm \bm{0.05}$ & $\bm{1.03} \pm \bm{0.006}$\\	
\bottomrule
\end{tabular}
%\ReduceSpaceUnderTable
\end{table}

\begin{table}[ttt]
\caption{Results of side-by-side (SxS) test with 95\% confidence intervals. Positive score means Method-A is preferred.}
\vspace{-6pt}
\label{tab:sxs_result}
\centering
\begin{tabular}{ c c | c }
\toprule
\textbf{Method-A} & \textbf{Method-B} & \textbf{SxS}\\	
\midrule
WaveGrad & PriorGrad & $0.165 \pm 0.061$\\
SpecGrad & WaveGrad  & $0.161 \pm 0.065$\\
SpecGrad & PriorGrad & $0.360 \pm 0.075$\\
\bottomrule
\end{tabular}
\ReduceSpaceUnderTable
\end{table}

The results for the noise schedules of WaveGrad and PriorGrad are shown in Tables \ref{tab:wg_schedule_result} and \ref{tab:pg_schedule_result}, respectively.
For all training and inference noise schedules, SpecGrad achieved the best scores for both subjective and objective metrics; SpecGrad provided better quality than WaveGrad and PriorGrad.
Since it worked well regardless of the choice of noise schedule, SpecGrad should be able to enhance the performance of other DDPM-based neural vocoders that use extended noise schedules~\cite{Bddm022,InferGrad2022}.

We additionally conducted a 7-scale (-3 to 3) side-by-side (SxS) preference test for PG-6 inference schedule. 
The results are shown in Table~\ref{tab:sxs_result}. 
This test also indicated that SpecGrad is better than WaveGrad and PriorGrad.
Note that PriorGrad was rated worse than WaveGrad in both MOS and SxS scores.
We found that artifacts generated by phase distortion of high frequency components were noticeable in PriorGrad.
Our experiments set the cutoff frequency of log-mel spectrogram to 12 kHz because the sampling frequency was 24 kHz, whereas that of the PriorGrad paper was 7.6 kHz~\cite{priorgrad}.
This difference can be a reason for the inconsistency between our and their experiments.
The unique property of SpecGrad is reduction of high frequency components according to the conditioning log-mel spectrogram (see Fig.~\ref{fig:concept}).
This feature can contribute to prevention of the artifacts because the amount of subtraction of the estimated noise ($\hat{\bm{\epsilon}}$ in Algorithm \ref{algo:sampling}) becomes smaller for higher frequency.
%, which should stabilize the training of the DNN and inference using it.

\vspace{-2pt}
\subsection{Evaluation as speech enhancement backend}
\label{sec:se_backend}

Since the diffusion PDF of SpecGrad depends on the conditioning log-mel spectrogram $\bm{c}$, its error might degrade the quality of output signals.
Therefore, we investigated the robustness of SpecGrad to the error of $\bm{c}$.
As a realistic scenario, we applied the DDPM-based neural vocoders as backends of an SE system.
WARP-Q and extended short-time objective intelligibility measure (ESTOI)~\cite{estoi} were used for objectively evaluating the speech quality and intelligibility, respectively.
Note that we used WARP-Q instead of the common metrics in speech enhancement~\cite{Maiti_waspaa_2019,Maiti_icassp_2020} such as the perceptual evaluation of speech quality (PESQ)~\cite{pesq}, CSIG, CBLK, and COVL~\cite{CsigCbakCovl} because they are designed for waveforms sampled at 16 kHz.

\vspace{2pt}
\noindent
\textbf{Model and training setup:}
We followed the SE scheme of parametric resynthesis~\cite{Maiti_waspaa_2019}, i.e., the frontend predicts the clean log-mel spectrogram from an observed noisy log-mel spectrogram, and then the backend generates a waveform given the predicted log-mel spectrogram.
For the SE frontend, we used a combination of a part of DF-Conformer~\cite{Koizumi_waspaa_2021} and Post-Net of Tacotron2~\cite{tacotron2};
an observed noisy log-mel spectrogram was enhanced by the mask predictor of DF-Conformer, and then its output was further cleaned up by the Post-Net.

We used the ``DF-Conformer-8'' model~\cite{Koizumi_waspaa_2021}, whose bottleneck feature dimension was adapted to that of the input log-mel spectrogram (i.e., 128), and Post-Net in Tacotron2~\cite{tacotron2} without modification.
The total number of their parameters was 7.6M.
This frontend was pretrained to minimize the sum of the mean-absolute- and mean-squared-error between the clean and predicted log-mel spectrograms at before and after Post-Net~\cite{tacotron2}.
The optimizer and batch size settings were the same as those of the training of SpecGrad, and we pretrained all models for 500k steps.
Then, the joint network was built by concatenating the pretrained frontend and a backend in Section \ref{sec:result} that was trained using WaveGrad noise schedule.
This joint network was finetuned for 500k steps to minimize the sum of the frontend loss and backend loss.
In the inference stage, WG-50 noise schedule was used for all models.

\vspace{2pt}
\noindent
\textbf{Dataset:} 
Training and test datasets were generated by contaminating the clean data used in Section \ref{sec:result} with reverberation and noise.
A room impulse response (RIR) for each sample was generated by a stochastic RIR generator using the image method~\cite{image_method}.
Its parameters were drawn from the following uniform distributions $\mathcal{U}$:
the distance between the source and microphone was $\mathcal{U}(0.5, 3.0)$ [cm], 
the length of one side of the square room was $\mathcal{U}(2.0, 10.0)$ [m], and 
the reflection ratio was $\mathcal{U}(0.5, 0.95)$. 
For the noise dataset, the TAU Urban Audio-Visual Scenes 2021 dataset~\cite{tau_2021_dataset} were used.
The average ratio of the energy of noise to the reverberated clean speech was 0 dB.
The average ESTOI and WARP-Q scores of the generated noisy samples were $51.9 \pm 1.1$ \% and $2.78 \pm 0.024$, respectively.

\vspace{2pt}
\noindent
\textbf{Results:} 
The results are shown in Table \ref{tab:sep_result}.
For both metrics, the proposed method outperformed WaveGrad and PriorGrad.
The differences of the scores between the proposed and conventional methods were smaller compared to those of the previous experiments, which should be because the frontend limited the maximum possible quality of the output signals (note that both ESTOI and WARP-Q measure the quality relative to the clean signals).
Even so, these results suggested that SpecGrad is robust to the error of the conditioning log-mel spectrogram, at least on the similar level with the conventional methods.

\begin{table}[ttt]
\caption{Results for speech enhancement experiment. ESTOI and WARP-Q scores with their 95\% confidence intervals.}
\vspace{-6pt}
\label{tab:sep_result}
\centering
\begin{tabular}{ c | c c }
\toprule
\textbf{Method} & \textbf{ESTOI [\%] ($\uparrow$)} & \textbf{WARP-Q ($\downarrow$)}\\	
\midrule
%WaveRNN & $77.9 \pm 0.7$ & $2.20 \pm 0.018$\\ 
WaveGrad & $82.7 \pm 0.7$ & $1.97 \pm 0.019$\\ 
PriorGrad & $81.9 \pm 0.7$ & $1.92 \pm 0.019$\\ 
SpecGrad & $\bm{83.6} \pm \bm{0.7}$ & $\bm{1.89} \pm \bm{0.019}$\\ 
\bottomrule
\end{tabular}
\ReduceSpaceUnderTable
\end{table}

% -------------------------------------------------------------------------
\section{Conclusion}
\label{sec:conclusion}

We proposed \textit{SpecGrad} that adapts the spectral envelope of diffusion noise based on the conditioning log-mel spectrogram.
We designed the decomposed covariance matrix $\bm{L}$ and its approximate inverse using the idea from T-F domain filtering.
This design allows us to use an FFT algorithm for computation of the matrix multiplication, which only adds a negligible extra computational cost compared to the forward computation of the conventional DDPM-based neural vocoders.
The experimental results showed that SpecGrad generated waveforms of higher quality than the conventional methods.
As SpecGrad performed well for different noise schedules, its combination with recently proposed extended noise schedules~\cite{Bddm022,InferGrad2022} is promising.

We believe SpecGrad could be further improved by incorporating more ideas from signal processing.
Some time-domain filters can be considered for the covariance matrix~\cite{Tokuda_2015,Tokuda_2016}.
The ideas from classic vocoders should be useful \cite{neural_source_filter,period_net} because SpecGrad can be viewed as a classic vocoder with noise excitation.
%Some prior information on phase and/or glottal-related events~\cite{Tokuda_2016} should also be helpful for improving phase at higher frequency.
These can be possible future work.

\newpage 

\bibliographystyle{IEEEtran}
\bibliography{mybib}

\end{document}